\documentstyle[prl,aps,preprint]{revtex}
\begin{document}
\draft
\title{One-way Functions In Reversible Computations}
\author{H. F. Chau\footnotemark ~and H.-K. Lo\footnotemark}
\address{
 School of Natural Sciences, Institute for Advanced Study, Olden Lane,
 Princeton, NJ 08540
}
\date{\today}
\preprint{IASSNS-HEP-95/47; quant-ph:9506012}
\footnotetext[1]{Present Address: Department of Physics, University of
Hong Kong, Pokfulam Road, Hong Kong e-mail: hfchau@hkusua.hku.hk}
\footnotetext[2]{Present Address: BRIMS, Hewlett-Packard Labs,
Filton Road, Stoke Gifford, Bristol BS12 6QZ, U. K. email:
hkl@hplb.hpl.hp.com}
\maketitle
\begin{abstract}
 One-way functions are used in modern crypto-systems as doortraps
 because their inverse functions are supposed to be difficult to compute.
 Nonetheless with the discovery of reversible computation, it seems
 that one may break a one-way function by running a reversible
 computer backward. Here, we argue that reversible computation alone poses
 no threat to the existence of one-way functions because of the
 generation of ``garbage bits'' during computations.
 Consequently, we prove a necessary and sufficient condition for a one-to-one
 function to be one-way in terms of the growth rate of the total number
 of possible garbage bit configurations with the input size.
\end{abstract}
\pacs{\noindent
 \begin{minipage}[t]{5.5in}\vspace{1ex}
  Physics and Astronomy Classification Numbers: 89.80.+h, 02.10.-v, 89.70.+c
  \vspace{1ex} \\
 \begin{tabular}{ll}
  AMS 1991 Subject Classification: & (Primary) 68Q05, 68Q15, 81P99 \\
  & (Secondary) 03D15, 94A60
 \end{tabular}
 \vspace{0.9ex} \\
 \begin{tabular}{ll}
  Keywords: & Garbage Bits, One-way Function, Quantum Computation, \\
  & Reversible Computation
 \end{tabular}
 \end{minipage}
}
 All presently known classical cryptographic schemes are based on the one-way
 function hypothesis. That is, there exist one-to-one functions that can be
 computed in times polynomial in the number of bits of their input arguments
 (and hence ``efficient'') while computation of their inverses cannot (and
 hence ``inefficient''). For example, the difficulty of eavesdropping in the
 well-known RSA public key cryptography is based on the observation that
 multiplying two numbers can be done efficiently while no known algorithm based
 on {\it classical} computers can factorize a composite number $N$ into primes
 in a time which is polynomial in $\log N$ \cite{Public_Key}. Thus, the
 existence of one-way functions is an important issue of crypto-security.
\par
 It has been shown that calculations can be done with reversible machines
 containing only reversible primitives
 \cite{Bennett:73,Toffoli:80,Fredkin:82,Ben}. One may be tempted by this
 reversibility to believe that one can invert a one-to-one function just by
 running a reversible computer backward. If this were true, the computation of
 the inverse function would be as efficient as the computation of the function
 itself. Such a naive argument seems to suggest that reversible computation
 rules out one-way functions and threatens the security of public key
 cryptography. In this article, however,
 we argue that reversible logic does not exclude the possibility of one-way
 functions. Their robustness relies on our ignorance of the values of the
 ``garbage bits'' in computations. The key issue is
 how the total number of possible garbage bit configurations
 produced in a reversible computation of a function grows with the
 input size. We define an efficient (in a time polynomial in the input
 size) reversible algorithm for computing a function to be {\it controllable}
 if the total number of garbage bit configurations is polynomial in the
 size of the input. It is then straight-forward to show that, given a
 one-to-one function $f$ that can be computed efficiently, a necessary and
 sufficient condition for it to be one-way is that it cannot be computed by
 any controllable algorithm.
\par
 Computation as it is currently carried out by electronic digital computers
 destroys information. For example, the so-called AND gate has two inputs and
 one output. When the output is $0$, we lose information because the input can
 be $(0,0)$, $(0,1)$ or $(1,0)$. Erasure of information is a dissipative
 process which costs energy. Incidentally, this observation leads to the first
 correct understanding of the Maxwell's demon \cite{Demon}.
\par
 However, Bennett \cite{Bennett:73,BL} has shown that all logic operations can
 be performed reversibly by adding extra redundancy in both the input and the
 output. Thus, in principle, one can build a computer without internal power
 dissipation. Consider the Toffoli gate (also known as controlled-controlled-NOT
 gate) \cite{Toffoli:80,Fredkin:82} which has
 three input and three output lines. The first two lines ($a$ and $b$) act as
 control and pass through the gate unchanged. The value of the third output
 line ($c_o$) depends on the third input line ($c_i$) and the control
 lines ($a$ and $b$), and is given by
\begin{equation}
 c_o = \left\{ \begin{array}{ll} c_i & \hspace{0.25in} \mbox{if~} a = 0
 \mbox{~or~} b = 0 \\ \mbox{NOT}(c_i) & \hspace{0.25in} \mbox{if~} a = b = 1
 \end{array} \right. \mbox{~.}
\end{equation}
 The Toffoli gate is clearly reversible: The input can be deduced from the
 output by running the latter through another Toffoli gate. Moreover, any logic
 operations can be implemented by an appropriate arrangement of the Toffoli
 gates. For example, by {\em presetting} the third input line to be zero, the
 third output bit will implement the logical AND operation on the first two
 input bits.
\par
 The above example illustrates two general features of reversible computations.
 First, to simulate an irreversible logic operation, one is required to preset
 some of the input bits at some particular (and fixed) values. Second,
 reversible systems produce not only what you want in the output (the logical
 ``AND'' between $a$ and $b$ in the third output bit), but also some ``garbage
 bits''. These garbage bits are, however, important. As noted before,
 the function AND is many-to-one. The garbage bits contain the information we
 need to run the computer in reverse. Actually, the Toffoli gate has shown to
 be able to perform universal computation in the sense that every function
 which is computable by the Turing machine is also computable (equally
 efficiently) by a reversible Turing machine \cite{Toffoli:80,Fredkin:82}.
 Moreover, we can confine the growth the garbage bits as follows
 \cite{Bennett:73,Feynman:86}:
\par\medskip\noindent
{\it Theorem~1:} (Bennett) The number of garbage bits in a reversible machine
 can be made equal to the number of input bits.
\par\medskip\noindent
{\it Proof:} We represent Bennett's algorithm schematically as
\begin{eqnarray}
 & & \mbox{INPUT} + \mbox{PRESET}_{\rm 1} + \mbox{PRESET}_{\rm 2} \nonumber \\
 & \longrightarrow & \mbox{OUTPUT} + \mbox{GARBAGE} + \mbox{PRESET}_{\rm 2}
 \nonumber \\ & \longrightarrow & \mbox{OUTPUT} + \mbox{GARBAGE} +
 \mbox{OUTPUT} \nonumber \\ & \longrightarrow & \mbox{INPUT} +
 \mbox{PRESET}_{\rm 1} + \mbox{OUTPUT} \mbox{~.}
\end{eqnarray}
 First we run the machine forward, giving us the $\mbox{OUTPUT}$ and the
 $\mbox{GARBAGE}$. Then we can copy down each of the output bits reversibly
 using a controlled-NOT gate \cite{Feynman:86}. Finally, we run the machine
 backward. In this way, $\mbox{PRESET}_{\rm 1}$ are the internal registers used
 as temporary storage whose values are unaltered after the process,
 $\mbox{PRESET}_{\rm 2}$ are the preset bits used by the controlled-NOT gates,
 and $\mbox{INPUT}$ in the {\em output} lines are regarded as ``garbage bits''.
\hfill$\Box$
\par\medskip\noindent
{\it Remark~1:} Any function $f$ that can be computed
 efficiently by a universal Turing machine can also be computed efficiently by
 a {\em reversible} Turing machine (with a possible slowdown by a constant
 factor)\cite{Bennett:73,Toffoli:80}. Now Theorem~1 shows that $f$ can be
 computed efficiently by an algorithm which produces the same number of garbage
 bits as the number of input bits. Thus, we can always design a reversible
 algorithm to calculate a given computable function $f$ with the number of
 garbage bits required at most equals the number of input bits.
\par\medskip\noindent
{\it Remark~2:} One must distinguish careful between internal registers
 ($\mbox{PRESET}_{\rm 1}$) and
 garbage bits. The internal registers are used only as temporary storage
 and they are unchanged at the end of a computation. On the contrary,
 the garbage bits take unknown values at the end of a computation. We remark
 that the number of internal registers needed for a reversible
 computation can be reduced by an elegant hierarchical ``pebbling argument''
 due to Bennett \cite{Bennett:89,Levine:90}. This space-efficient
 simulation is obtained by breaking the original computation into
 {\it segments} and then doing and undoing these segments in a hierarchical
 manner. Moreover, the increase in running time is insignificant. However, this
 pebbling argument does not reduce the number of garbage bits.
\par\medskip
 In some cases, it may be possible to further reduce the number of garbage
 bits. If the function that we are evaluating is one-to-one, the desired output
 already contains all the information needed to deduce the input. No garbage
 bit is needed to account for the missing information. (The computer may still
 use some internal registers as scratch space during a calculation. We only
 need to make sure that those internal registers are restored to their initial
 values at the end of the calculation.) Given a function $f$, this machine will
 take the input $x$ to $y = f(x)$ without garbage bits; i.e., $\mbox{INPUT}
 \longrightarrow \mbox{OUTPUT}$. The computation of a one-to-one computable
 function $f$ without garbage bits was discussed by Bennett \cite{Bennett:73}:
\par\medskip\noindent
{\it Theorem~2:} (Bennett) Let $f$ be an invertible function. If $f$ and
 $f^{-1}$ are both computable, then there exits a reversible algorithm to
 compute $f$ without producing any garbage bits.
\par\medskip\noindent
{\it Proof:} Since every function computable using a Turing machine is also
 computable using a reversible Turing machine, we may assume the existence of
 reversible algorithms in computing $f$ and $f^{-1}$, each of them may produce
 some garbage bits. We now construct a new reversible algorithm which sends
 $\mbox{INPUT}$ to $\mbox{OUTPUT}$ without any garbage bits as shown below:
\begin{eqnarray}
 & & \mbox{INPUT} + \mbox{PRESET}_{\rm 1} + \mbox{PRESET}_{\rm 2} \nonumber \\
 & \longrightarrow & \mbox{OUTPUT} + \mbox{GARBAGE}_{\rm 1} +
 \mbox{PRESET}_{\rm 2} \nonumber \\ & \longrightarrow & \mbox{OUTPUT} +
 \mbox{GARBAGE}_{\rm 1} + \mbox{OUTPUT} \nonumber \\ & \longrightarrow &
 \mbox{INPUT} + \mbox{PRESET}_{\rm 1} + \mbox{OUTPUT} \nonumber \\ &
 \longrightarrow & \mbox{INPUT} +\mbox{GARBAGE}_{\rm 2} + \mbox{INPUT}
 \nonumber \\ & \longrightarrow & \mbox{INPUT} + \mbox{GARBAGE}_{\rm 2} +
 \mbox{PRESET}_{\rm 2} \nonumber \\ & \longrightarrow & \mbox{OUTPUT} +
 \mbox{PRESET}_{\rm 1} + \mbox{PRESET}_{\rm 2} \mbox{~.}
\end{eqnarray}
 This corresponds to running the $f$ machine forward, (reversibly) copying the
 output down using a number of controlled-NOT gates, then running the $f$
 machine backward. After that we run the $f^{-1}$ machine forward and then
 reversibly erase the input (again by means of a series of controlled-NOT
 gates \cite{Feynman:86}). Finally, we run the $f^{-1}$ machine backward. Note
 that $\mbox{PRESET}_{\rm 1}$ and $\mbox{PRESET}_{\rm 2}$ can be regarded as
 internal registers providing scratch space during the calculation.
\hfill$\Box$.
\par\medskip\noindent
{\it Remark~3:} Suppose we have efficient algorithms to compute both $f$ and
 $f^{-1}$, then it is easy to verify that the above ``zero garbage-bit
 algorithm'' to compute $f$ is also efficient.
\par\medskip
 Reversible computation gives rise to a paradox: a reversible machine seems to
 given a short cut to compute the inverse of a function. By running the machine
 backward, it takes $y$ into $f^{-1} (y) = x$. Thus, one may be tempted to
 believe that the computations of $f$ and $f^{-1}$ are equally efficient. If
 this were true, there would be no one-way functions and all classical
 cryptographic systems would be in danger. In view of the widespread usage of
 public key crypto-systems, it is crucial for us to resolve this paradox. (This
 question is brought forward again \cite{Proc} after the recent polynomial time
 factorization algorithm using quantum computer by Shor \cite{Shor}. And
 cryptography by quantum mechanical means may be the only secure method
 \cite{Proc,Bennett:92,Jozsa:94,Schumacher:95}. We shall return to this point
 later in this article.)
\par
 The resolution of the above paradox lies in the fact that in general there is
 no {\em efficient} way to reduce the garbage (without an efficient way of
 computing the inverse function and we shall elaborate on
 this point in the proof of Theorem~3). The value of each garbage bit may be 0
 or 1 depending on the input. However, the values taken by the various garbage
 bits may well be correlated so that when there are $k$ garbage bits, the total
 number of all possible configurations of the garbage bits is in general less
 than or equal to $2^k$. For a given $\mbox{OUTPUT}$, there is only one such
 combination which is correct in the sense that the appropriate input
 ($\mbox{INPUT}$) and preset bits ($\mbox{PRESET}$) can be obtained by running
 the machine backward. That is, $\mbox{OUTPUT} + \mbox{GARBAGE} \longrightarrow
 \mbox{INPUT} + \mbox{PRESET}$. (If the function is many-to-one, then there are
 multiple possible garbage bits combinations which are ``correct'' in general.
 Each of them corresponds to a pre-image of the function.) To efficiently
 compute the inverse function, we must be able to control the growth of the
 number of possible garbage bit configurations. This leads us to the following
 definition.
\par\medskip\noindent
{\it Definition~1:} An {\it efficient} reversible algorithm to compute $f$
 is said to be {\em controllable} if and only if all the potential values
 of the garbage bits are known and the total number of possible garbage bit
 configurations scales like a polynomial function of the number of input bits.
\par\medskip\noindent
{\it Remark~4:} The number of different possible garbage bit configurations
 depends greatly on the reversible algorithm. The trick by Bennett
 \cite{Bennett:73} that we have described in Theorem~1 above is in general
 {\em not} effective enough to control the growth of the number of possible
 garbage bit configurations. The pebbling argument discussed in Remark~2
 is not useful neither.
\par\medskip
 With the concept of a controllable algorithm in mind, we can prove a theorem
 which measures the difficulty in computing an inverse function.
\par\medskip\noindent
{\it Theorem~3:} Given a one-to-one function $f$ which can be computed
 efficiently, then the following statements are equivalent:
\par\indent
(a) $f$ is not a one-way function;
\par\indent
(b) there exists an efficient algorithm to compute $f(x)$ without generating
 any garbage bits;
\par\indent
(c) there exists an efficient algorithm to compute $f(x)$ with the number of
 garbage bits that scales as the logarithm of the input bit length; and
\par\indent
(d) $f(x)$ can be computed by a controllable algorithm.
\par\medskip\noindent
{\it Proof:} (a) $\Rightarrow$ (b) is just a direct application of Theorem~2
 and Remark~3. Besides, it is easy to see that (b) $\Rightarrow$ (c)
 $\Rightarrow$ (d). To show the equivalence of the above four statements, it
 remains to prove that (d) $\Rightarrow$ (a).
\par
 We fix the $\mbox{OUTPUT}$ to be the configuration that represents $y$. Then
 $x = f^{-1} (y)$ is obtained in the following way. First, we set the garbage
 bits to one of the possible configurations before running the machine
 backward. It is clear that if the values of the preset bits agree with the
 ones we have used when we run the machine forward, then the input bit
 configuration is in the pre-image of $\mbox{OUTPUT}$ under the function $f$.
 Since $f$ is one-to-one, we can conclude that the input bit configuration
 corresponds to $x = f^{-1} (y)$. Thus, we test if the values of the preset
 bits agree with the ones we have used when we run the machine forward. If they
 agree, then reading out the $\mbox{INPUT}$ will give us the value of $x$, and
 we are done. If not, we run the machine forward again, and then (reversibly)
 replace the garbage bits by another possible configuration (by means of a
 table). This process is repeated until the preset bits agree with the ones we
 have used. Since all the potential garbage bit configurations are known and
 their total number is at most a polynomial function of the input bit length,
 the above reversible algorithm is efficient. Thus, $f$ is not a one-way
 function. This completes the proof.
\hfill$\Box$
\par\medskip\noindent
{\it Remark~5:} An efficient algorithm for $f^{-1}$ is obtained provided that
 we can find an efficient way to ``control'' the growth of the number of
 garbage bits. Besides, Theorem~3 tells us that finding an efficient way to
 control the growth a garbage bits is as difficult as finding an efficient way
 to compute the inverse function itself. In other words, Theorem~3
 states that, given a one-to-one function that can be computed efficiently,
 a necessary and sufficient condition for it to be one-way is that it cannot be
 computed by any controllable algorithm. Therefore, reversible computation
 does not rule out one-way function. It only makes its definition
 more precise.
\par\medskip\noindent
{\it Remark~6:} Even if the garbage bit configuration is known, the method use
 in Theorem~3 to compute the inverse function is only as efficient as computing
 the function $f$ itself. Other algorithms (if any) have to be used if we
 demand an algorithm of computing $f^{-1}$ which is more efficient than
 computing $f$.
\par\medskip
 It is instructive to give an example of a function that can be computed
 by a controllable algorithm.
\par\medskip\noindent
{\it Example~1:} Consider the reversible algorithm of sending $x$ to $x + 1
 \mbox{~mod~} 2^n$ as shown in Figure~1. Obviously, the algorithm is efficient.
 The number $x$ is encoded in the binary
 representation $(a_{n-1} a_{n-2} \cdots a_1 a_0)$, and $c_i$ ($i = 1,2,\ldots
 ,n-2$) are the garbage bits use to keep track of the carries. Since $a_0$ can
 be used as a carry for the least significant digit, there is no need to
 invoke $c_0$. It is straight forward to
 see that the possible values for the output garbage bits are $(c_{n-2} c_{n-3}
 \cdots c_1) = (00\cdots 00), (00\cdots 01), (00\cdots 11), \ldots, (11\cdots
 11)$. Thus the total number of possible garbage bit configurations grows
 linearly with the number of input bits $n$ and hence the algorithm is
 controllable.
\par\medskip\noindent
{\it Remark~7:} Similarly, it can be shown that the reversible algorithm of
 adding two numbers modulo $2^n$ (i.e., $a + b + \mbox{PRESET} \longrightarrow
 (a+b) + b + \mbox{GARBAGE}$) requires $\mbox{O}(n^2)$ garbage bits.
 The total number of possible garbage bits configurations grows as
 $\mbox{O}(n^2)$, and hence the algorithm is controllable. Thus, once $a+b$ and
 $b$ are given, $a$ can be found efficiently.
\par\medskip
 It is difficult to give examples of functions that are efficiently
 computable {\it only} by algorithms that are {\it not} controllable
 because showing their existence would be equivalent to proving the
 (still un-proven) one-way function hypothesis.
\par
 We remark that reversible computation is quite common in microscopic chemical
 reactions. DNA copying in nature may be regarded as a special form of
 reversible computation. Extensive works have been done recently to explore the
 power of massive parallelism in DNA computation \cite{Adleman,Lipton}.
\par
 In recent years, a new quantum theory of computation has been developed
 \cite{Feynman:86,Feynman:82,Deutsch_1}. A quantum computer can follow many
 distinct computation paths simultaneously and produce a final output depending
 on the interference of all of the paths. In particular, the Shor's efficient
 quantum algorithm to factor large composite numbers \cite{Shor} challenges the
 existence of one-way functions and threatens the security of public key
 crypto-systems. It has been suggested that perfect security can be achieved
 only by quantum cryptography \cite{Proc,Bennett:92,Jozsa:94,Schumacher:95}.
 Unlike classical cryptography which is based on the unproven one-way function
 hypothesis, quantum cryptography is unbreakable because it relies on the
 ``no-cloning theorem'' of non-orthogonal states in quantum mechanics.
 Nevertheless, quantum
 cryptography requires quantum coherence which is technologically difficult to
 maintain over long distances. For example, a quantum particle transmitted from
 the Voyager spacecraft back to the Earth will undoubtedly lose its coherence
 due to interactions with solar wind on its way. In the foreseeable future, the
 widespread usage of classical cryptography is likely to continue. It is,
 therefore, important to examine the robustness of the one-way function
 hypothesis. And we have shown in this article that this hypothesis is
 consistent with the classical reversible computers.
\par
 A simple remark is in order. For a quantum computer to speed up
 computation, it is crucial for the various computational paths to
 interfere with each other. In other words, new {\it quantum} algorithms
 are needed to exploit the massive ``quantum parallelism''. If we are to
 run a {\it classical} algorithm (devised for classical reversible computers)
 in a quantum computer, it is well-known that there will be no speed up.
 (Compare with Remark~6)
 For such an algorithm, the best thing we can do is to prepare some
 superpositions of garbage bits configurations and
 run the quantum machine backward. Then we make a measurement on the preset
 bits to see if they  agree with the ones we will use when we want to run the
 machine forward. On average, the number of trials we have to perform before an
 agreement is made scales exponentially with the number of garbage bits. Thus,
 this method is inefficient in general.
\acknowledgements{This work is supported by the DOE grant DE-FG02-90ER40542.}

\begin{figure}
 \caption{An reversible algorithm for adding unity modulo $2^5$. Algorithm for
  adding unity modulo $2^n$ can be constructed in a similar way. The number
  is encoded as $(a_4 a_3 a_2 a_1 a_0)$. The $c_i$ are the garbage bits, which
  is used to keep track of the carry. Standard reversible logic notations are
  used where $\oplus$ denotes negation, and $\bullet$ denotes a controlling
  bit. For example, the first and the second operations in this figure are the
  controlled-controlled-NOT and controlled-NOT gates respectively.}
\end{figure}
\newpage
\begin{center} {\bf Biographical Sketches} \end{center}
\par
 Hoi Fung Chau was born in 1968. He received his B.Sc. in mathematics and
 physics in 1989, and Ph.D. in physics in 1992, both from the University of
 Hong Kong. He was an astrophysics postdoc in University of Illinois
 at Urbana-Champaign from 1992 to 1994, and a member of the Institute for
 Advanced Study in Princeton NJ from 1994 to 1996. Currently, he is an
 assistant professor in physics in the University of Hong Kong.
 His research interests include astrophysics of
 neutron stars, cellular automata, quantum computation, self-organized
 criticality, and statistical physics.
\par
 Hoi-Kwong Lo is currently a postdoctoral research consultant at
 the Basic Research Institute in Mathematical Sciences, Hewlett-Packard
 Laboratories, Bristol, U. K. He received his B.A. in mathematics from
 Trinity College, Cambridge University, U.K. in 1989;
 M.S. and Ph.D. in physics from Caltech, Pasadena in 1991 and
 1994 respectively. In 1994-96, he was a member of the Institute of
 Advanced Study, Princeton NJ.
 His current research interest is quantum information and
 computation.
\end{document}